\documentclass[a4paper,11pt]{article}
\usepackage{pos}
\usepackage{amsfonts,amsmath}
\usepackage{graphicx}
\usepackage{bm}
\usepackage{color}

\title{%
\vspace{-20mm}%
\begin{flushright}%
\begin{minipage}{0.26\textwidth}
\normalsize%
KEK Preprint 2025-11 \\ CHIBA-EP-269 
\end{minipage}	
\end{flushright}
\vspace{10mm}
Further numerical evidences for the gauge-independent separation between Confinement 
and Higgs phases in lattice SU(2) gauge theory with a scalar field in the fundamental representation
}

\ShortTitle{%
Further numerical evidences for the gauge-independent separation $\cdots$
}

\author*[a,b]{Akihiro Shibata}
\author[c]{Kei-Ichi Kondo}

\affiliation[a]{%
Computing Research Center, High Energy Acceelerator Research Organization (KEK),\\
Oho 1-1, Tsukuba 305-0801, Japan}
\affiliation[b]{%
SOKENDAI (The Graduate University for Advanced Studies), 
Oho 1-1, Tsukuba 305-0801, Japan}
\affiliation[c]{%
Department of Physics, Chiba University, 
1-33 Yayoi-cho, Chiba 260,  Japan}

\emailAdd{$^{a,b}$akihiro.shibata@kek.jp}
\emailAdd{$^c$kondok@faculty.chiba-u.jp}

\abstract{%
In the lattice gauge-scalar model with a single scalar field in the fundamental representation of 
the gauge group $SU(2)$\,, we have quite recently found 
that there exists a gauge-independent transition line separating the Confinement and Higgs phases 
without contradicting the well-known Osterwalder-Seiler-Fradkin-Shenker analyticity 
theorem between the two phases by performing numerical simulations without any gauge fixing. 
This was achieved by examining the correlation between the original fundamental scalar field 
and the so-called color-direction field constructed from the gauge field 
through the gauge-covariant decomposition due originally to Cho-Duan-Ge-Shabanov and Faddeev-Niemi.

In this presentation, we give further numerical evidence for the gauge-independent separation 
between the Confinement and Higgs phases in the above model to establish their physical origin. 
For this purpose, we investigate the  separation line  precisely. 
We also investigate the contributions of magnetic monopoles 
to examine their role in confinement 
from the viewpoint of the dual superconductor picture.
}

\FullConference{The XVIth Quark Confinement and the Hadron Spectrum Conference (QCHSC24)\\
 19-24 August, 2024\\
 Cairns Convention Centre, Cairns, Queensland, Australia\\}

\DeclareMathOperator{\tr}{tr}
\DeclareMathOperator{\re}{Re}

\begin{document}
\maketitle

\section{Introduction}

We investigate the gauge-scalar model on the lattice to clarify the mechanism 
of Confinement in the Yang-Mills theory in the presence of matter fields. 
We also investigate the non-perturbative characterization of 
the Brout-Englert-Higgs (BEH) mechanism \cite{Higgs1}, 
providing the gauge field with the mass in a gauge-independent way (without gauge fixing). 
However,  it is impossible to realize the conventional BEH mechanism
on the lattice unless the gauge fixing condition is imposed 
since gauge non-invariant operators have vanishing vacuum expectation 
value on the lattice without gauge fixing due to the Elitzur theorem \cite{Elitzur75}.
This difficulty can be avoided by using the 
\textit{gauge-independent description of the BEH mechanism} proposed recently 
by one of the authors, 
which needs neither the spontaneous breaking of gauge symmetry nor
the non-vanishing vacuum expectation value of the scalar field  \cite{Kondo16,Kondo18}.
Therefore, we introduce the gauge-independent description of the BEH mechanism 
on the lattice and study the Higgs mechanism in a gauge-invariant way.

As for the gauge-scalar model with radially fixed scalar field (no Higgs mode)
which transforms according to the \textit{fundamental} representation of the gauge group $SU(2)$,
we have quite recently found that there exists a gauge-independent transition line 
separating the Confinement and Higgs phases without contradicting the well-known
Osterwalder-Seiler-Fradkin-Shenker analyticity theorem \cite{FradkinShenker79,Ostewlder78}
between the two phases by performing numerical simulations without any gauge fixing \cite{Ikeda23}.
On the other hand, for the gauge-scalar model with a radially-fixed scalar field 
in the \textit{adjoint} representation of the gauge group $SU(2)$,
Brower et al. have shown that the Confinement and Higgs phases are completely 
separated into two different phases by a continuous transition line in the unitary gauge \cite{Brower82}.
However, we have recently found a new transition line that divides completely the Confinement
phase into two parts without gauge fixing \cite{ShibataKondo23}.

This presentation gives further numerical evidence for the gauge-independent separation
between the Confinement and Higgs phases in the above model to establish its physical origin. 
For this purpose, 
we accumulated the data for an extensive set of parameters ($\beta$, $\gamma$) 
using the method in \cite{Ikeda23, ShibataKondo23}  with higher statistics. 
We further investigate the magnetic monopole 
in order to clarify the physical characteristics of each phase
from the view of the dual superconductor picture.

\section{SU(2) lattice gauge-scalar model and gauge covariant decomposition}
We introduce the lattice SU(2) gauge-scalar model with a single scalar field in the fundamental representation of the gauge group where the radial degrees of freedom of the scalar field is fixed 
(no Higgs modes) \cite{Ikeda23}.
The action of this model with the gauge coupling constant $\beta$ and the scalar coupling constant 
$\gamma$ is given in the standard way by
\begin{align}
S[U,\hat{\Theta}]
&= 	\frac{\beta}{2} \sum_{x,\mu>\nu} 
	 \re \tr \left( 
	   \mathbf{1} - U_{x,\mu} U_{x+\hat{\mu},\nu} U_{x+\hat{\nu},\mu}^{\dagger} U_{x,\nu}^{\dagger} 
	  \right) 
  +  \frac{\gamma}{2} \sum_{x,\mu} 
     \re \tr \left( 
        \mathbf{1} - {\hat{\Theta}}_x^{\dagger} U_{x,\mu} {\hat{\Theta}}_{x+\hat{\mu}} 
     \right) ,
\label{action1}
\end{align}
where $U_{x,\mu} \in \mathrm{SU(2)}$ is a (group-valued) gauge variable on a link $\langle x,\mu \rangle$, 
and ${\hat{\Theta}}_x \in \mathrm{SU(2)}$ is a (matrix-valued) scalar variable 
in the fundamental representation on a site $x$ which obeys the unit-length (or radially fixed) condition: 
${\hat{\Theta}}^{\dagger}_x {\hat{\Theta}}_x = \bm{1} ={\hat{\Theta}}_x {\hat{\Theta}}^{\dagger}_x$ \,.
This action is invariant under the local $\mathrm{SU(2)}_\mathrm{local}$ gauge transformation 
and the global $\mathrm{SU(2)}_\mathrm{global}$ transformation for the link variable $U_{x,\mu}$ 
and the site variable ${\hat{\Theta}}_x$:
$U_{x,\mu} \mapsto U_{x,\mu}^{\prime} = {\Omega}_x U_{x,\mu} {\Omega}_{x+\mu}^{\dagger}$ and 
$\hat{\Theta}_x \mapsto \hat{\Theta}_x^{\prime} = {\Omega}_x \hat{\Theta}_x \Gamma$ for 
${\Omega}_x \in \mathrm{SU(2)}_\mathrm{local}$\,, 
$\Gamma \in \mathrm{SU(2)}_\mathrm{global}$\,.

In our investigations, the color-direction field defined shortly plays the key role. 
This new field was introduced in the framework of change of field variables 
which is originally based on the gauge-covariant decomposition \cite{CFNdecomp07,Exactdecomp09}
of the gauge field due to 
Cho-Duan-Ge-Shabanov\cite{Cho80,Duan-Ge79,Shabanov99} and Faddeev-Niemi\cite{FN98}. 
(see \cite{KKSS15} for a review.)

The \textit{color-direction field} on the lattice is a (Lie-algebra valued) site variable:
$\bm{n}_x := n_x^A {\sigma}^A \in \mathrm{su(2)-u(1)}$ $(A=1,2,3)$\ with 
the unit length $\bm{n}_x \cdot \bm{n}_x = 1$, where ${\sigma}^A$ are the Pauli matrices. 
We require the transformation property of the color-direction field $\bm{n}_x$ as
$\bm{n}_x \mapsto \bm{n}_x^{\prime} = {\Omega}_x \bm{n}_x {\Omega}_x^{\dagger}$\,.

For a given gauge field configuration $\{ U_{x,\mu} \}$, 
we determine the color-direction field configuration $\{ \bm{n}_x \}$ 
(as the unique configuration up to the global color rotation) by minimizing the so-called 
\textit{reduction functional} $F_\mathrm{red} [\bm{n} ; U]$ under the gauge transformations:
\begin{align}
	F_\mathrm{red} [\{ \bm{n}\}  ; \{ U\}]
	&:= \sum_{x,\mu} \frac{1}{4} \tr 
	\left[ {\left( D_{\mu} [U] \bm{n}_x \right)}^{\dagger} \left( D_{\mu} [U] \bm{n}_x \right) \right] 
	= \sum_{x,\mu} \frac{1}{2} \tr \left( \bm{1} - \bm{n}_x U_{x,\mu} \bm{n}_{x+\hat{\mu}} U_{x,\mu}^{\dagger} \right) \,.
	\label{red}
\end{align}
In this way, a set of color-direction field configurations $\{ \bm{n}_x \}$ is obtained as the (implicit)
functional of the original link variables $\{ U_{x,\mu} \}$, which is written symbolically as 
\begin{align}
	\bm{n}^*= \underset{\bm{n}}{\text{argmin}} \, F_\mathrm{red} [\{ \bm{n}\}  ; \{ U\}].
	\label{redc2}
\end{align}
This construction shows the nonlocal nature of the color-direction field. 

By way of the color-direction field, the original link variable $U_{x,\mu} \in \mathrm{SU(2)}$ is gauge-covariantly decomposable into the product of two field variables $X_{x,\mu}, V_{x,\mu} \in \mathrm{SU(2)}$: $U_{x,\mu} = X_{x,\mu} V_{x,\mu}$\,. 
We require that $V_{x,\mu}$ has 
the transformation law in the same form as the original link variable $U_{x,\mu}$ 
and that $X_{x,\mu}$ has the transformation law in the same form as the site variable $\bm{n}_x$:
\begin{align}
	V_{x,\mu} \mapsto V_{x,\mu}^{\prime} = {\Omega}_x V_{x,\mu} {\Omega}_{x+\hat{\mu}}^{\dagger} \, ,
	\quad 
	X_{x,\mu} \mapsto X_{x,\mu}^{\prime} = {\Omega}_x X_{x,\mu} {\Omega}_x^{\dagger} \, .
\end{align}
This decomposition is uniquely determined by solving the \textit{defining equations} simultaneously 
(once the color-direction field is given):
\begin{align}
	D_{\mu} [V] \bm{n}_x := V_{x,\mu} \bm{n}_{x+\hat{\mu}} - \bm{n}_x V_{x,\mu} = 0 \, , 
	\quad 
	\tr \left( \bm{n}_x X_{x,\mu} \right) = 0 \, ,
\end{align}
where $D_{\mu} [V]$ denotes the covariant derivative in the adjoint representation.
Indeed, the exact solution is obtained in the following form \cite{Exactdecomp09} :
\begin{align}
	V_{x,\mu} &= \tilde{V}_{x,\mu} / \sqrt{\frac{1}{2}\tr \left( \tilde{V}_{x,\mu}^{\dagger} \tilde{V}_{x,\mu} \right)} \,, \quad 
	\tilde{V}_{x,\mu} := U_{x,\mu} + \bm{n}_x U_{x,\mu} \bm{n}_{x+\hat{\mu}} \, ,\quad 
	X_{x,\mu} = U_{x,\mu} V_{x,\mu}^{\dagger} \, .
	\label{eq:XV}
\end{align}

By introducing the color-direction field, we obtain the deformed theory 
in which the expectation value of an operator $\mathscr{O}$ including the color-direction field 
is calculated according to 
\begin{align}
	& \langle \mathscr{O}[U,\hat{\Theta}, \bm{n} ]  \rangle 
	= \frac{1}{Z} \int \mathcal{D} U \mathcal{D} \hat{\Theta} e^{-S[U,\hat{\Theta}]} \int \mathcal{D} \bm{n} \ \bm{\delta}(\bm{n} -\bm{n}^*)  \mathscr{O}[U,\hat{\Theta},\bm{n} ] ,
	\label{exvev2}
\end{align}
where $\mathcal{D} \bm{n}=\prod_x d\bm{n}_x$ is the invariant measure for the color-direction field 
and $\bm{\delta}(\bm{n} -\bm{n}^*)$ is the Dirac delta function which plays the role of replacing $\bm{n}$ 
by $\bm{n}^*$ determined by (\ref{redc2}).

It should be remarked that these new variables have been successfully used to understand confinement 
based on the dual superconductor picture. For example, it has been shown in the pure gauge theory 
without the matter field that the restricted field $V$ gives the dominant part for quark confinement, 
while the remaining field $X$ corresponds to the massive modes and decouples in the low-energy region. 
This gives the gauge-independent version of the Abelian dominance observed in the maximal Abelian gauge. 
See \cite{KKSS15} for more details and more applications of this reformulation of the gauge theory. 

\section{Lattice results}
We perform the numerical simulation on the $16^{4}$ lattice with the periodic boundary condition.
Link variables $\{U_{x,\mu}\}$ and scalar fields $\{\Theta_{x}\}$ are updated alternately 
by using the HMC (Hamiltonian Monte Carlo) algorithm with integral interval 
$\Delta\tau=1$ without gauge fixing.
After the thermalization of 5000 sweeps, we store 3000 configurations every five sweeps.

We search for a separation line 
that separates the phases or distinguishes the physical origin of confinement
by measuring the expectation value 
$\left\langle \mathcal{O} \right\rangle$ of a chosen operator $\mathcal{O}$ 
by changing $\gamma$ (or $\beta$) along the $\beta=\text{const.}$ (or $\gamma=\text{const.}$) line.
We identify the separation line 
by using singular property such as bends, steps, jumps, and gaps 
observed in the graph of the $\left\langle \mathcal{O} \right\rangle$ plots
or peaks in the graph of susceptibility $\chi(\mathcal{O})$ plots.

\subsection{Thermodynamic phase transition}
\begin{figure}\centering
\includegraphics[width=0.40\textwidth] {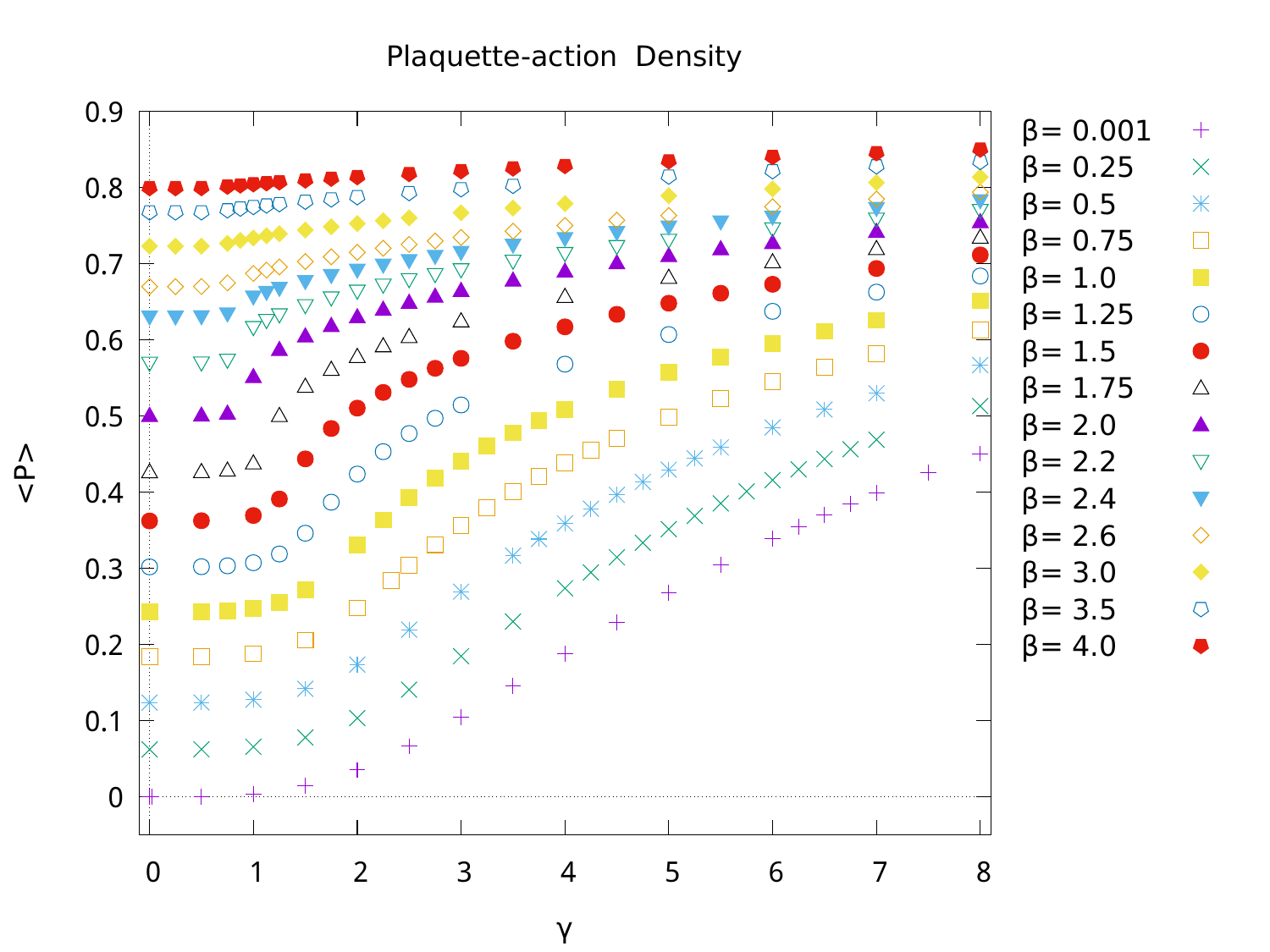}%
\hspace{5mm}   		
\includegraphics[width=0.40\textwidth] {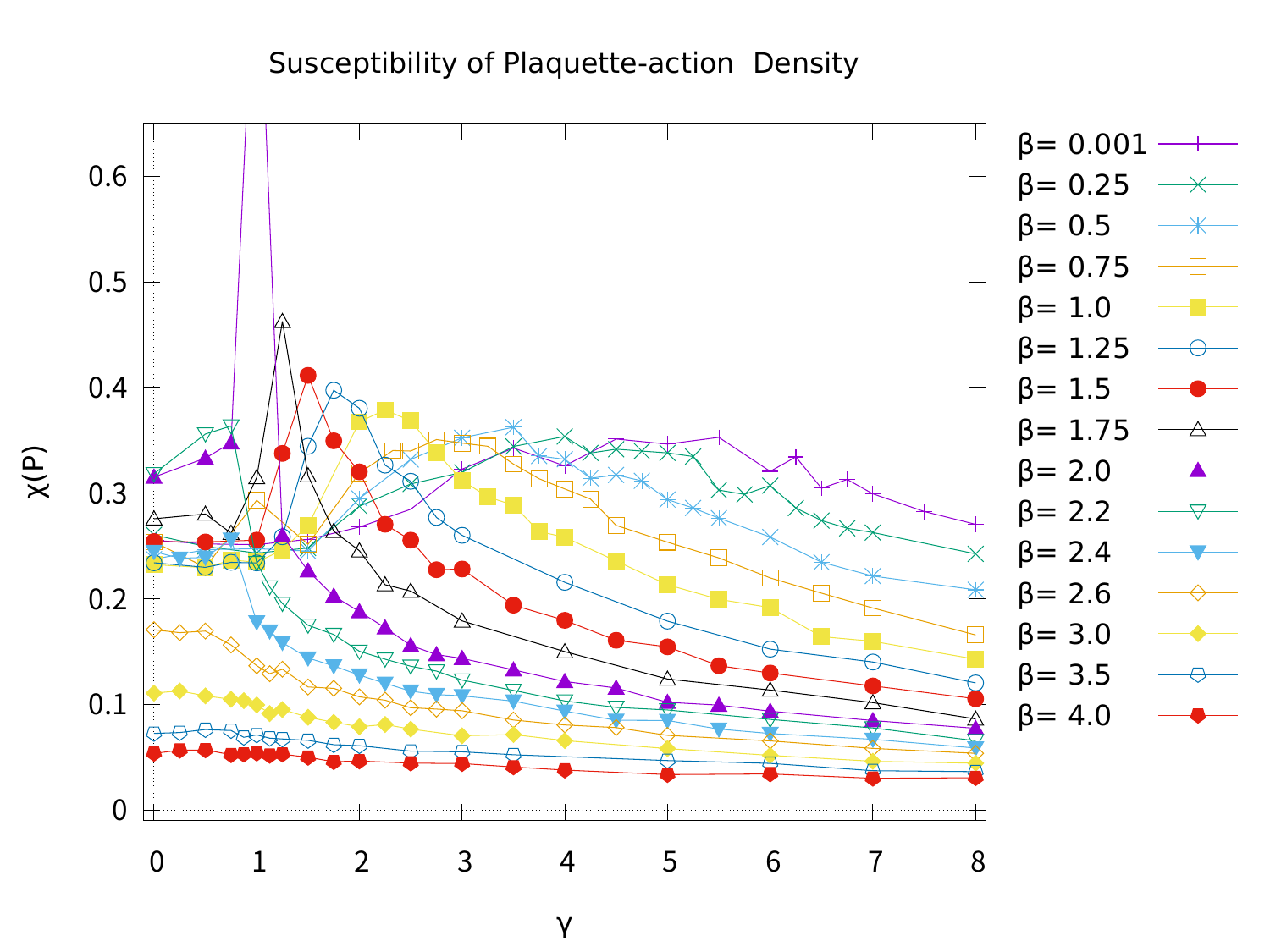}%
\caption{%
Left: Plots of $\left\langle P \right\rangle$ for various $\beta=\text{const.}$
Right: Plots of   $\chi(P) $ for various $\beta=\text{const.}$
Lines in the plots are the eye guides for $\beta=\text{const.}$
}
\label{fig:P}
\end{figure}%
\begin{figure}\centering
\includegraphics[width=0.40\textwidth] {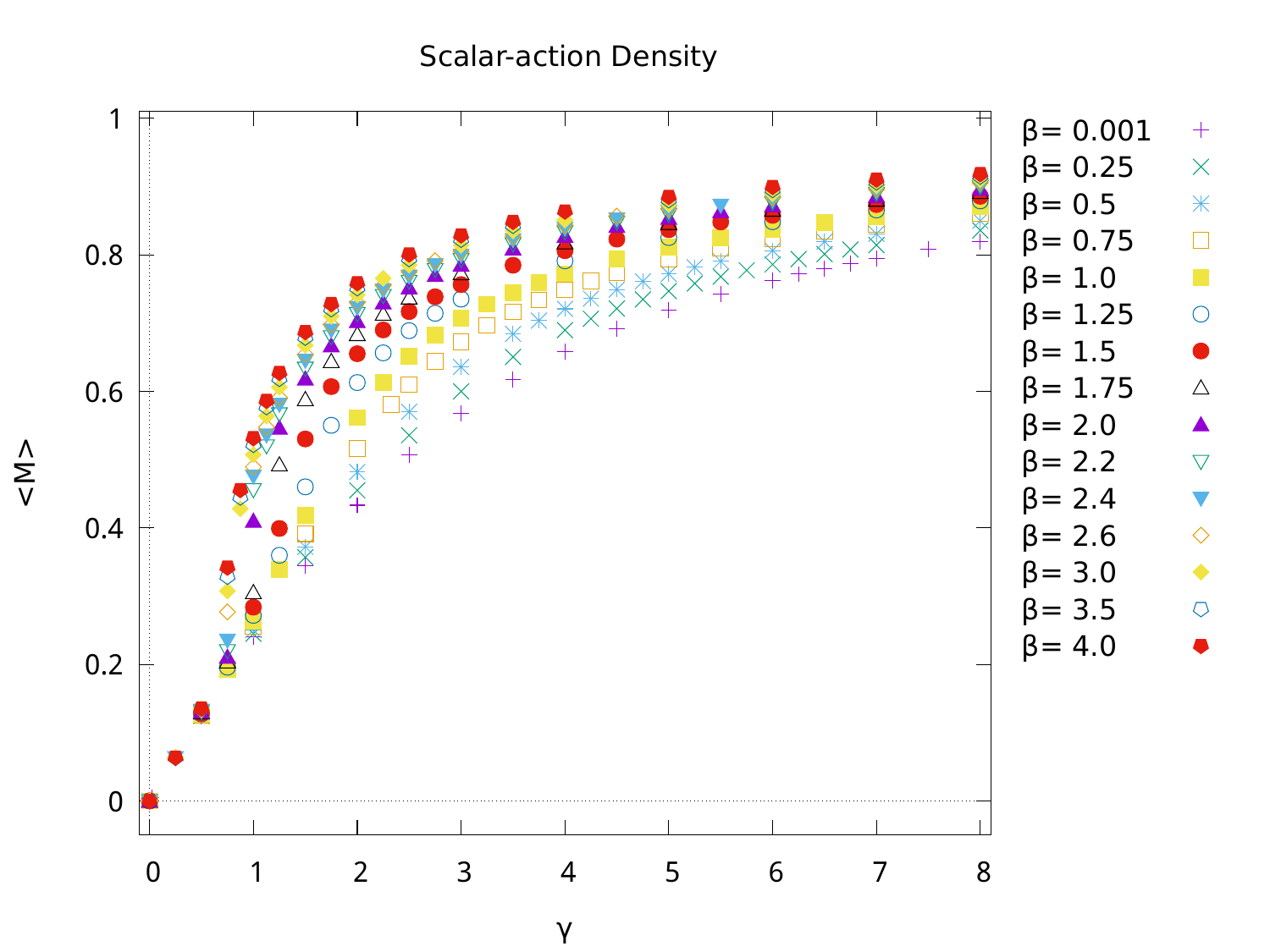}%
\hspace{5mm}	
\includegraphics[width=0.40\textwidth] {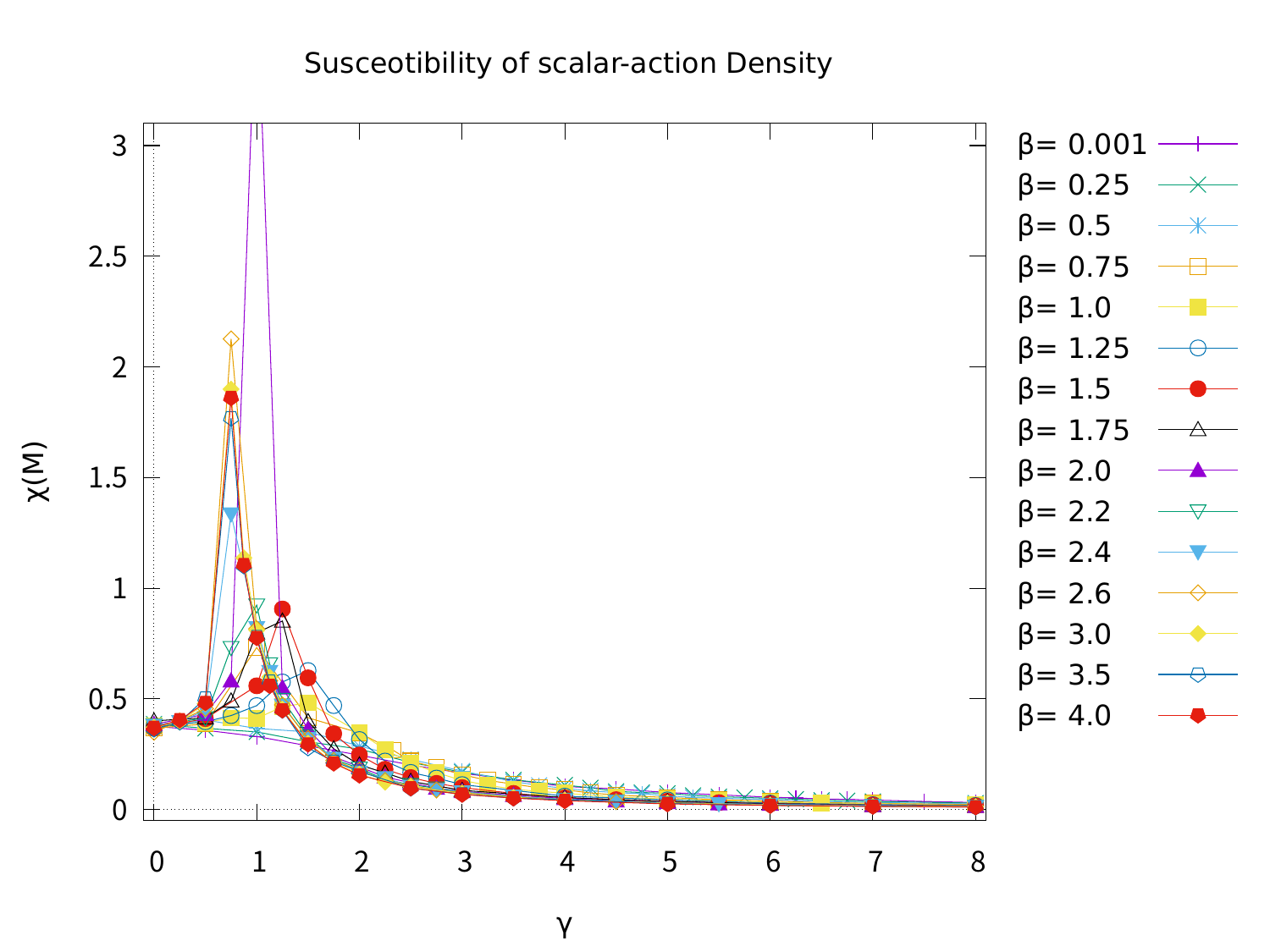}%
\caption{%
 Left: Plots of  $\left\langle M \right\rangle$ for various  $\beta=\text{const.}$
 Right: Plots of $\chi(M)$ for various  $\beta=\text{const.}$
}
\label{fig:M}
\end{figure}
We reexamine the separation line obtained from the action density 
with high statistics for an extensive set of parameter ($\beta$, $\gamma$)
as in the previous work \cite{Ikeda23}:
\begin{align}
P &=\frac{1}{6N_{\text{site}}}\sum_{x}
 \sum_{\mu<\nu}\frac{1}{2}\mathrm{tr} 
 \left( 
  U_{x,\mu}U_{x+\hat{\mu},\nu}U_{x+\hat{\nu},\mu}^{\dag}U_{x,\nu}^{\dag}
 \right) \,,\quad 
 \chi(P) = ( 6N_{\text{site}} ) \left\{
 \left\langle P^2 \right\rangle - \left\langle P \right\rangle^{2} 
 \right\} \,,    
 \label{eq:P} \\
M &=\frac{1}{4N_{\text{site}}}\sum_{x}\sum_{\mu}\frac{1}{2} Re\mathrm{tr}\left(
\Theta_{x}^{\dag} U_{x,\mu} \Theta_{x+\hat{\mu}} \right) \,, \quad
\chi(M) = (4N_{\text{site}}) 
  \left\{
  \left\langle M^2 \right\rangle - \left\langle M \right\rangle^{2}
  \right\} \,,
\label{eq:M}%
\end{align}
where $\left\langle {\mathcal O} \right\rangle$ represents the average of the operator ${\mathcal O}$ over configurations. 
In this analysis, we focus on the susceptibility to determine the separation line.

Figure \ref{fig:P} shows the result of measurements for the plaquette density (\ref{eq:P}).
The left panel shows the plot of $\left\langle P \right\rangle$ for various $\beta=\text{const.}$\,,
and the right panel shows the plot of the susceptibility $\chi(P)$ for various $\beta=\text{const.}$ 
There exist bends or gaps in the $\left\langle P \right\rangle$
and peaks in the $\chi(P)$  which correspond to the  separation line.
Note that there are two types of peaks for $\chi(P)$\,: 
a narrow and sharp peaks observed in $\beta > \beta_c \simeq 0.75$, 
and a broad and gradual peaks observed in  $\beta < \beta_c$.
In the same way, Fig.\ref{fig:M} shows the result of measurements for the scalar density (\ref{eq:M}).
The left panel shows the plot of  $\left\langle M \right\rangle$ for various $\beta=\text{const.}$, 
and the right panel the plot of susceptibility $\chi(M)$ for various $\beta=\text{const.}$

The left panel of Fig.\ref{fig:phase} shows the phase diagram determined 
from the action density.
The dark blue plots represent the separation line determined from both the average of action densities,
$\left\langle P \right\rangle$,  $\left\langle M \right\rangle$, and their susceptibilities, $\chi(P)$, $\chi(M)$.
This separation line could be the first order phase transition line
and disappears in the region $\beta < \beta_c$, 
which is  pointed out by
the Osterwalder-Seiler-Fradkin-Shenker analyticity theorem \cite{FradkinShenker79,Ostewlder78}.
On the other hand, the orange plots in $\beta < \beta_c$ represent 
the separation line suggested only by the susceptibility $\chi(P)$.
This separation line could be the second order phase transition or the cross over. 
However, those peaks are not narrow and sharp but broad and gradual.

In addition, there exists the other separation line 
obtained only from $\chi(P)$
(the orange plots in the region $0 \le \gamma < 1$ and $2 < \beta <2.5$)
that separates between the scaling and non-scaling region, 
which corresponds to the "critical" point already discovered 
in the pure $SU(2)$ Yang-Mills theory ($\gamma=0$) in \cite{BhanotCreutz81, Creutz82}.

\begin{figure}[hbt] \centering
\includegraphics[width=0.41\textwidth] {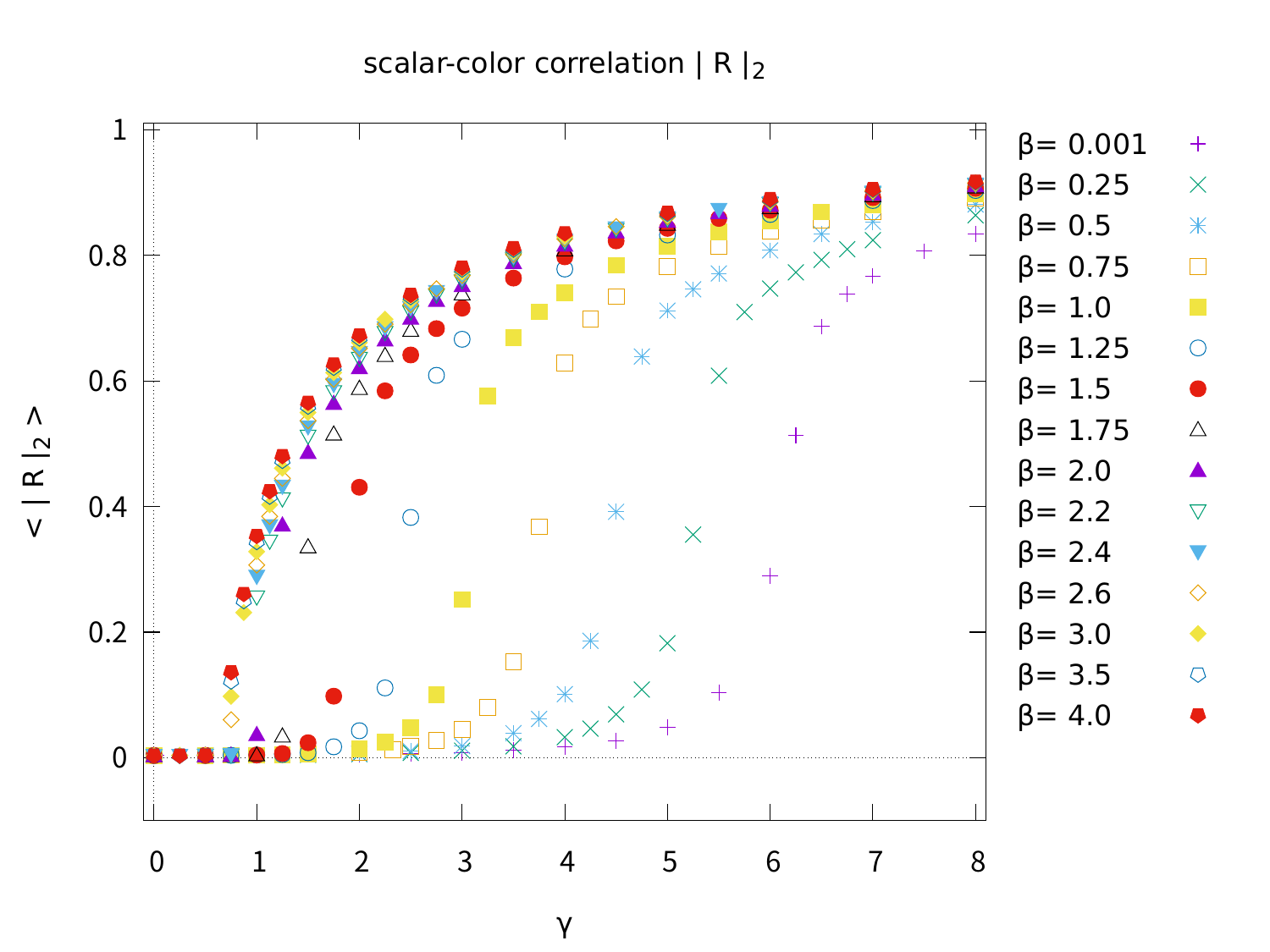}%
\hspace{5mm}
\includegraphics[width=0.41\textwidth] {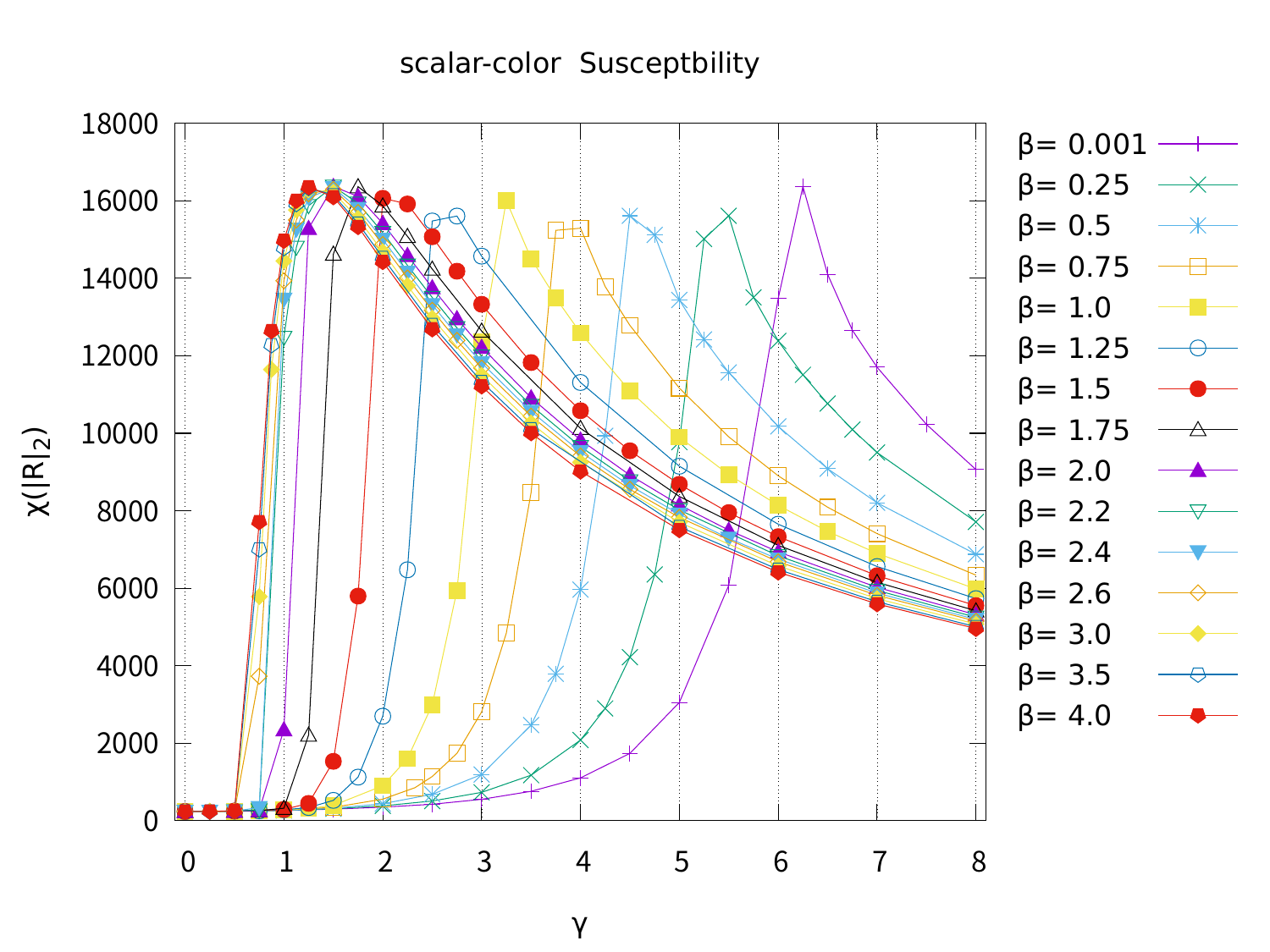}%
\caption{%
Left: Plots of $\left\langle \left\vert \bm{R} \right\vert_{2} \right\rangle$
for various  $\beta=\text{const.}$
Right: Plots $\chi(\left\vert \bm{R} \right\vert_{2})$
for various $\beta=\text{const.}$
}
\label{fig:R}
\end{figure}		
\subsection{Scalar-color correlation}
In order to  reexamine the correlation between the scalar field 
and the color-direction field, as in the previous work \cite{Ikeda23}:
\begin{align}
& \bm{R}=\frac{1}{N_{\text{site}} }\sum_{x} \Theta_{x}^{\dag} \mathbf{n}_{x} \Theta_{x}\,
\quad  
\left\langle \left\vert \bm{R} \right\vert_{2} \right\rangle
\,, \quad  
\chi(\left\vert \bm{R} \right\vert_{2})=(4N_{\text{site}})
  \left\{
	\left\langle\  \left\vert \bm{R} \right\vert_{2}^2 \right\rangle
	- \left\langle\  \left\vert \bm{R} \right\vert_{2} \right\rangle^2
  \right\} \,,
\end{align}
where $\left\vert \bm{R} \right\vert_{2} $ represents
the 2-norm defined by 
$ \left\vert \bm{R} \right\vert_{2} %
= \sqrt{ |R_1|^2 + |R_2|^2 + |R_3|^2}$
with $R_A = \text{tr}\left(\bm{R} \sigma^{A}\right)$.

The left panel of Fig.\ref{fig:R} shows the plots of 	
$\left\langle \left\vert \bm{R} \right\vert_{2} \right\rangle$ v.s. $\gamma$ 
for various $\beta=\text{const.}$ 
We observe bends or gaps for all the $\beta=\text{const.}$, which correspond
to the separation line.
The right panel of Fig.\ref{fig:R} shows the plots of 
the susceptibility $\chi(\left\vert \bm{R} \right\vert_{2})$ v.s. $\gamma$ 
for various $\beta=\text{const.}$\,, 
and we observe sharp peaks for all the $\beta=\text{const.}$

\begin{figure}[hbt] \centering
\includegraphics[width=0.41\textwidth]{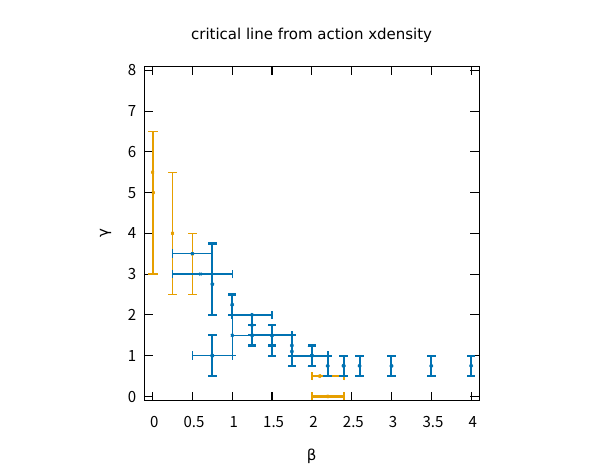}
\includegraphics[width=0.41\textwidth]{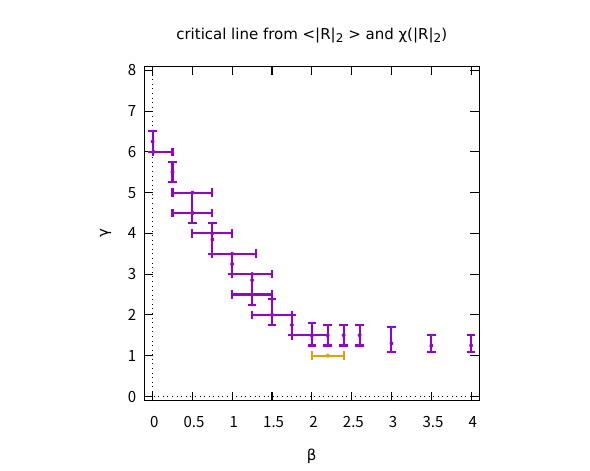}
\caption{%
Left: The phase transition line determined from action density,
$\left\langle P \right\rangle$, $\chi(P)$,  
$\left\langle M \right\rangle$, and $\chi(M)$. 
The orange plots represent that obtained only from $\chi(P)$.
Right: The separation line determined from the scalar-color correlation
$\left\langle\left\vert \bm{R} \right\vert_{2} \right\rangle$, 
$\chi(\left\vert \bm{R} \right\vert_{2})$.
}
\label{fig:phase}
\end{figure}

The right panel of Fig.\ref{fig:phase} shows the separation line obtained 
from the scalar-color correlation. We determine the location of the phase separation line
by using peaks of the susceptibility $\chi(\left\vert \bm{R} \right\vert_{2})$ plots
rather than the bents or jumps 
in the $\left\langle \left\vert \bm{R} \right\vert_{2} \right\rangle$ plots,
which causes its location to shift toward 
a larger $\gamma$ compared to the previous work \cite{Ikeda23}.
This suggest the separation between 
the Confinement and Higgs phases completely for all the region.

\subsection{Contribution of the magnetic monopole}
Moreover,  we investigate the contribution of the magnetic monopole 
to clarify the physical origin of the separation of the Confinement and Higgs phases
in view of the dual superconductor picture \cite{dualsuper}
where magnetic monopoles play the dominant role in confinement.

We can define the magnetic monopole, $k_{x,\mu}$ in a gauge-independent (gauge-invariant) way
through the gauge-covariant decomposition \cite{KKSS15}:
\begin{align}
 & F(x)_{\mu\nu} 
 := \arg_{F}\text{tr}
     \left\{ (\bm{1}+\bm{n}_{x}) 
      V_{x,\mu}V_{x+\hat{\mu}, \nu} V^{\dag}_{x+\hat{\nu},\mu}V^{\dag}_{x,\nu}
    \right\} \,, \nonumber \\
 & k_{x,\mu} := \frac{1}{2} \epsilon^{\mu \nu \alpha \beta} 
	\left(F(x+\hat{\nu})_{\alpha,\beta} - F(x)_{\alpha\beta} \right\}
   =: 2\pi m_{x,\mu} \,, \quad m_{x,\mu} = 0, \pm 1, \pm2, \cdots\,,
\end{align}
where $V_{x,\mu}$ represents the restricted field obtained from the gauge-covariant decomposition (\ref{eq:XV}), 
and $\bm{n}_{x}$ the color-direction field determined by (\ref{redc2}).
This magnetic monopole takes the integer value and satisfies the current conservation law, i.e., 
$\partial_{\mu} k^{x,\mu} = \sum_{\mu} ( k_{x+\hat{\mu},\mu}-k_{x,\mu})=0$. 
Therefore, we define the magnetic-monopole-charge density:
\begin{align}
 \rho_{k} := \frac{1}{4 N_{\text{site}}} \sum_{x,\mu} | m_{x,\mu} |\,.
\end{align}

\begin{figure}[bt] \centering
\includegraphics[width=0.37\textwidth] {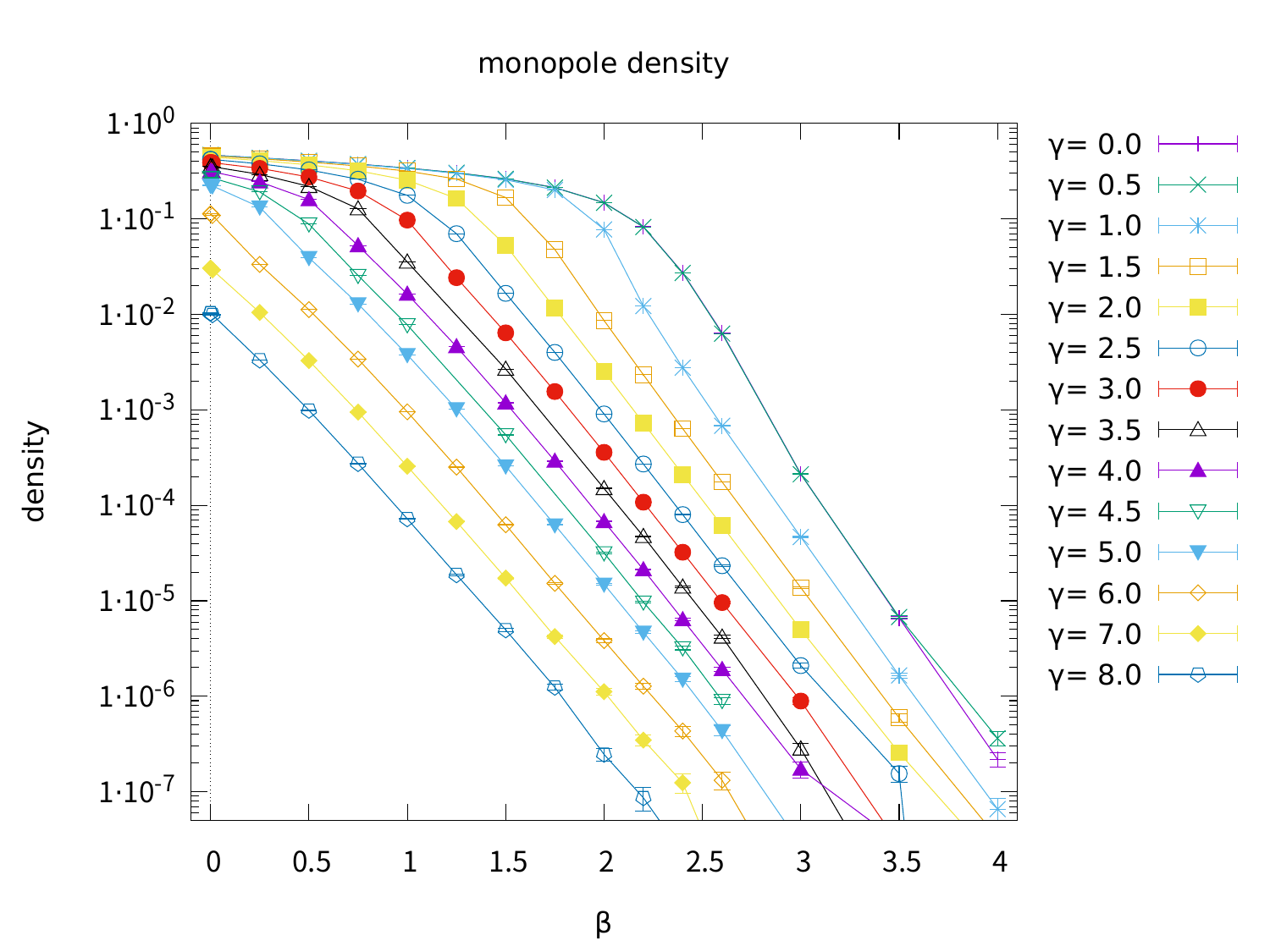}%
\hspace{6mm}
\includegraphics[width=0.37\textwidth] {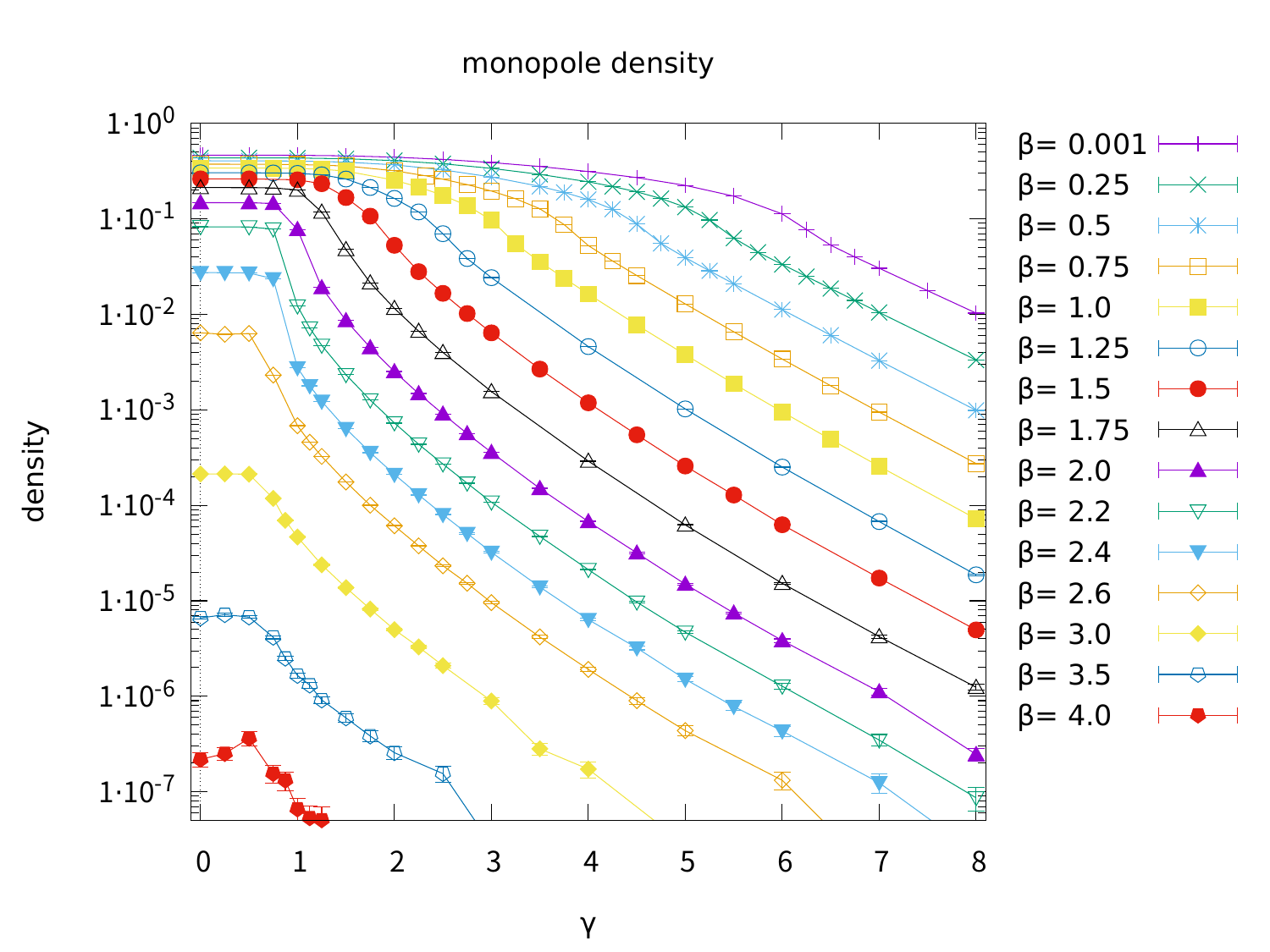}%
\caption{
Left: Plots of $\left\langle \rho_{k}\right\rangle$ v.s. $\beta$  for various $\gamma=\text{const.}$
Right: Plots of $\left\langle \rho_{k}\right\rangle$ v.s. $\gamma$ for various  $\beta=\text{const.}$
The lines in the plots are eye guide with $\beta=\text{const.}$ or $\gamma=\text{const.}$  
}
\label{fig:monopole}
\end{figure}

Figure \ref{fig:monopole} shows the plot of 
the magnetic-monopole-charge density $\left\langle\rho_{k}\right\rangle$.
The left panel shows the plot of $\left\langle\rho_{k}\right\rangle$
along various $\gamma=\text{const.}$ lines.
$\left\langle\rho_{k}\right\rangle$ decreases as $\beta$ increases
and there exist no singular points in the plots.
Note that
we observe very small $\left\langle\rho_{k}\right\rangle$ for large $\beta$,
which is measured not in the physical unit but in the lattice one. 
This is because the lattice spacing in the physical unit is a decreasing function of $\beta$,
and for large $\beta$ the physical volume becomes small against the lattice with fixed size.

The right panel of Fig.\ref{fig:monopole} shows $\left\langle\rho_{k}\right\rangle$ along 
various $\beta=\text{const.}$ lines. 
In the region $\gamma < \gamma_c(\beta)$,
where $\gamma_c(\beta)$ represents the separation line in the right panel of Fig.\ref{fig:phase}),
the magnetic-monopoles-charge density 
$\left\langle\rho_{k}\right\rangle$ is rich and constant along the $\beta=\text{const.}$ line.
It suggests that the magnetic monopole plays a dominant role in this region, 
which we call Confinement phase.
In contrast, we observe less or vanishing magnetic monopoles 
in the region $\gamma > \gamma_c(\beta)$.
In the region $\beta > \beta_c \simeq 0.75$, 
we observe bends or steps  at $\gamma$ being the phase separation line, 
$\gamma=\gamma_c(\beta)$, and the magnetic monopoles disappear for  $\gamma > \gamma_c(\beta)$.
Therefore, the gluons gain mass due to the BEH mechanism and are confined.
We call this region the Higgs phase.
While, in the region $\beta < \beta_c$
there exist no more bends or steps for $\left\langle\rho_{k}\right\rangle$,  
and $\left\langle\rho_{k}\right\rangle$ decreases rather smoothly as $\gamma$ increase.
The separation line $\gamma_c(\beta)$ could not be a phase transition line 
but be the boundary where the two physical origins switch over continuously.
\section{Summary}
We have reexamined the $SU(2)$ gauge-scalar model with the scalar field in the fundamental representation
to obtain further numerical evidence for the gauge-independent separation 
between the Confinement and Higgs phases based on the method in the previous work \cite{Ikeda23,ShibataKondo23}.
We have confirmed the thermodynamic phase transition, 
which is consistent with the Osterwalder-Seiler-Fradkin-Shenker analyticity theorem.
We have also reexamined the separation line 
that separates the Confinement and Higgs phases
based on the covariant decomposition of the gauge field and confirmed it.
Although it could not be the thermodynamic phase transition line, 
it could be the separation line that distinguishes the physical origin of the confinement. 

Moreover, we have investigated the magnetic-monopole-charge density 
to clarify the physical origin of the separation line.
We have confirmed that in the Confinement phase
the magnetic monopole plays a dominant role in confinement,
while in the Higgs phase the magnetic monopoles disappear and the Yang-Mills field
acquires the mass through the BEH mechanism and is confined.
\subsection*{Acknowledgments}
This work was supported by Grant-in-Aid for Scientific Research, JSPS KAKENHI Grant
Number (C) No.23K03406. 
The numerical simulation was supported in part by the Multidisciplinary
Cooperative Research Program in CCS, University of Tsukuba. 


\end{document}